# Communicative competencies and the structuration of expectations:
## The creative tension between Habermas' critical theory and Luhmann's social systems theory

Loet Leydesdorff

Amsterdam School of Communications Research (ASCoR), University of Amsterdam,
Kloveniersburgwal 48, 1012 CX Amsterdam, The Netherlands;
loet@leydesdorff.net; http://www.leydesdorff.net

**Abstract**

I elaborate on the tension between Luhmann's social systems theory and Habermas' theory of communicative action, and argue that this tension can be resolved by focusing on language as the interhuman medium of the communication which enables us to develop symbolically generalized media of communication such as truth, love, power, etc. Following Luhmann, the layers of *self-organization* among the differently codified subsystems of communication versus *organization* of meaning at contingent interfaces can analytically be distinguished as compatible, yet empirically researchable alternatives to Habermas' distinction between "system" and "lifeworld." Mediation by a facilitator can then be considered as a special case of organizing historically contingent translations among the evolutionarily developing fluxes of intentions and expectations. Accordingly, I suggest modifying Giddens' terminology into "a theory of *the structuration of expectations*."

## Introduction

In their ambitious—empirical, programmatic, and philosophical—paper entitled "Between Chaos and Entropy: Community of Inquiry from a Systems Perspective," Nadia Stoyanova Kennedy and David Kennedy try to bridge the gap between Luhmann's (1995) social systems theory and Habermas' (1992) "ideal speech situation" by using the concept of a community of inquiry in which learning can be facilitated. These authors argue that the autopoietic inquiring system can be offered as an exemplar of Habermas' "ideal speech situation." In another context, I have argued that the difference between these two opposing positions can be resolved from the perspective of a sociological theory of communication (Leydesdorff, 2000, 2001).

From this perspective, the human agent can be appreciated not only as a necessary condition for communication—using Maturana's (1978) mechanism of "structural coupling"—but because of an additional coupling in language as also contingent in terms of expectations and intentions (Luhmann, 2002, at pp. 175 and 182). Not only are agents "structurated" by systems as aggregates of action—Giddens' (1979, 1984) "duality of structure" (cf. Leydesdorff, 1993)—but additionally their expectations are culturally "structurated" by "horizons of meaning" (Husserl, 1929). More recently, I elaborated this elsewhere into a structuration theory of expectations (Leydesdorff, 2010a). It seems to me that the two rich resources of semantics—critical theory and social systems theory—can be recombined into this structuration theory of expectations without losing empirical grounding or theoretical perspective.

Since the debates between Habermas and Luhmann during the 1970s (e.g., Habermas & Luhmann, 1971) both positions have been developed further. Habermas (1981) elaborated his theory of the ideal speech situation into a theory of *communicative action,* and Luhmann ([1984], 1995) absorbed, but modified Maturana's (1978; cf. Maturana & Varela, 1980 and 1984) concept of *autopoiesis* (self-organization) into a sociological framework (Parsons, 1968). In the further elaboration of social systems theory, Luhmann made several major steps on which I will now first expand in order to return thereafter to the relevance to Habermas' humanistic contribution with its emphasis on language. The focus on communication and discourse allows for integration into, among other things, the model of the community of inquiry as proposed by Stoyanova Kennedy & Kennedy (2010).

**"Meaning" and "life" as different *autopoieses***

The first major step by Luhmann (1986) was to absorb Maturana & Varela's (1980) theory of *autopoiesis* as a specification of the mechanism for emergence in recursive and non-linear interactions. Whereas Maturana & Varela (1984) studied "life" as the emergent result of exchanges among molecules and molecular structures at the biological level, the mechanism of *autopoiesis* can also be abstracted from this material substrate, formalized, and applied to other domains (Leydesdorff, 2001; Rosen, 1985). Luhmann (1990) had already proposed as a contribution to the discussion with Habermas that "meaning" and its dynamics should be considered as the proper domain of sociology: how is the "interhuman" system coordinated by communication? Luhmann's answer is that this system is specific in communicating meaning, and not uncertainty (that is, Shannon-type information; cf. Leydesdorff, 2002).

As against Maturana & Varela, who identified system layers uniquely in terms of *what* is communicated (Mason, 1991), Luhmann (1986) proposed considering *two* systems able to process meaning in a co-evolution: individual consciousness (that is, the psychological system) and interhuman communication (that is, the social system). These two systems precondition each other in a "structural coupling," but additionally "interpenetrate" each other because they have access to each other's substance reflexively. This reflection can be carried by language as an achievement of cultural evolution (Luhmann, 2002). Language additionally enables us to develop symbolically generalized media of communication (Parsons, 1963a and b; 1968) such as love, truth, power, etc. The specific codes of communication in each of these media shape "horizons of meaning" emerging from and reproduced by the communications among us. The communications can be considered as local instantiations of interactions among these horizons (Giddens, 1979).

Two forms of differentiation are distinguished: (*i*) functional differentiation between the symbolically generalized media of communication enables us to process more complexity than when all communication has to be integrated into a "self" as a single axis, and (*ii*) systems differentiation among three levels: interactions, organization of the communication in instantiations, and self-organization of the communication in terms of function systems. The symbolically generalized media of communication can be considered as performative in the sense that they enable us to process complexity, while organized media (printing, electronic) support organization in the instantiation by materializing the layer of messaging. The self-organization of meaning, however, remains "intangible" because it emerges and therefore cannot be observed directly (Luhmann, 1995, at p. 165). The informed reader will recognize in this latter position Husserl's (1935/6) critique of the positive sciences as a possible—or better: impossible—model for the social sciences.

The interactions among us generate variation for the social system. A variation can also be considered as probabilistic entropy or uncertainty, that is, Shannon-type information. Meaning emerges from *relating* bits of information, for example, in configurations. The configurations can be recognized reflexively as specific organizations of meaning in communications. When organization prevails, stratification and control can be expected. The self-organization, that is, the abandonment of control and its replacement with freedom(s), allows us to add another layer of meaning processing reflexively: functional differentiation of the symbolically generalized codes of communication operates globally; not locally in historically situated organizations.

Historically, this tension was first expressed as universal human values institutionalized into civil liberties (such as, the freedom of speech and religion), but from the perspective of hindsight these values can be recognized not as transcendental, but as contingently shaped in a continuous refinement of the communication at the level of society. The codes of the communication enable us to distinguish among contributions, noise, and transgression. Their functional differentiation enriches our repertoire.

For example, while one can negotiate about the price of a commodity at a street market, one can pay the price without further negotiations in a store. Money enables us to generalize the transaction symbolically. After the invention of coins, it took centuries to develop the notion of banknotes as a symbolic generalization. Credit cards and electronic fund transfers enable us to transmit larger amounts of money faster and more globally than one could ever exchange on a local market. These media thus make us performative.

It is neither a value (such as "the pursuit of happiness" as a human right) nor a Greek god (like Hermes or Mercury) that does this job, but the communication proceeds at the level of the social system so that the problems involved in this globalization can be solved incrementally. Once a solution is achieved, however, the result can click—resonate (Simon, 1969 and 1973a; Smolensky, 1986)—and reshape the system from which it was constructed because of the superior competencies embedded in it culturally. Control emerges and tends to invert the order from bottom-up to top-down.

Note that the resulting mechanism is not material, but remains an order of expectations that incurs on our acting in a *second* contingency, namely the contingency of our intentions. The "double contingency" (Parsons, 1968; Parsons & Shills, 1951, at pp. 14 ff.; Vanderstraeten, 2002) can be decomposed into a first contingency of our material lives and practices, and a second one of our reflexive expectations and intentions. These reflexive expectations can be attuned in communities that organize the meanings that one attributes to possible events and to each other, but they can also reach beyond the existing formats and resonate symbolically.

Stoyanova Kennedy & Kennedy (2010) are right that this introduces a dialectics between organized constraints and enabling in the communication *versus* liberty and self-organization of the communication. The materialization in the instantiations can be considered as a retention mechanism of the unfolding of horizons of meaning which analytically remain orthogonal and therefore incommensurable, but which have yet to be traded-off in each interhuman encounter.

This model assumes that all symbolically generalized dimensions of interhuman communication are present in all interhuman communication, but that acculturation is precisely the competence to perform selectively in these terms (Foucault, 1966). Thus, what can be said in parliament cannot be said in court without potentially disturbing a symbolic order. Economists may argue about "shortages of energy" or an energy crisis, while a physicist will know that energy is a conserved entity of which there can be no shortage. Discourses are differently codified as specific organizations of meaning guided by symbolic generalizations that may remain unspoken in restricted discourses, but can be elaborated further if necessary (Bernstein, 1971; Coser, 1975).

**The turn towards contingent idealism**

In the above I have used Luhmann's theorizing as a source of inspiration, but reached beyond it by sociologizing this theory into a structuration theory of expectations. Luhmann's (e.g., 1997) theory, in my opinion, remained too constrained by the meta-biological discourse of systems theory, with its focus on observers and on binary units of information which allow only for "ON/OFF". For example, he emphasized at several places that the codes operate in terms of binary values (e.g., true/false for the code of science), while the order of expectations, in my opinion, necessarily proceeds in terms of uncertainties. In scholarly discourse, for example, statements can no longer be considered unambiguously as true or false; heuristics (truth-finding) and puzzle-solving prevail in evolving discourses (Kuhn, 1962; Rorty, 1992; Simon, 1973b).

Unlike systems which process molecules, meaning-processing systems are not closed, but may *tend towards* closure. Closure remains an expectation, while the realization requires a trade-off at interfaces which *organize* the meaning processing. Under the pressure of cyberneticians (e.g., Von Foerster, 1982 and 1993), Luhmann during the 1990s increasingly adopted the metaphor of an observer, and thus the theorizing lost its relevance for the sociological enterprise, which analyzes in terms not of observations but of expectations (Leydesdorff, 2006a). (Of course, these expectations can be tested against carefully designed empirical observations.)

Thus, not only the systems under study became "closed," but also the theoretical system itself became "closured" into a new paradigm celebrated in the German philosophical tradition, that is, with an internally oriented self-understanding. However, Giddens (1976, at pp. 142 ff.) already critiqued the biological metaphor implied in Kuhn's (1962) notion of closure of the paradigm as follows:

> The process of learning a paradigm or language-game as the expression of a form of life is also a process of learning what that paradigm is not: that is to say, learning to mediate it with other, rejected, alternatives, by contrast to which the claims of the paradigm in question are clarified. (at p. 144).

He added in the introduction to the *Constitution of Society* (1984, at p. xxxvii):

> There can be no doubt about the sophistication and importance of the work of some authors currently endeavouring to develop Parsons's work in novel ways, particularly, Luhmann and Habermas. But I think it as necessary to repudiate the newer versions of Parsonianism as I do the longer established versions of non-Parsonian structural sociology.

However, his own alternative—the structuration theory of action—turns to practices and action which are eventually modeled individually. This actor is not monadically encapsulated like a Robinson Crusoe (Habermas, 1986, at p. 378), but embedded in discourses. However, the speech act remains an individual act, although the agent can perhaps also be considered in a social role as an institutional or principal agent (Giddens, 1981).

Habermas (1986) voiced the metaphor of a monadically encapsulated consciousness (à la Robinson Crusoe) in a critique of Luhmann because the focus on communication (the operation of the social system) as distinguished from consciousness (the operation of the human mind) models the two systems as monads which disturb one another in a structural coupling, but are no longer able to communicate meaningfully among themselves. As Habermas (1987, at p. 381) formulated:

> No common denominator can be built up among different psychic systems, unless it be an autocatalytically emergent social systems, which is immediately locked again within its own systemic perspectives and draws back into its own egocentric observational standpoints: "This capacity to process information may suffice for the few aspects relevant to interaction (among mutually observing, self-referential systems). They remain separate, they do not fuse, they do not understand one another better than before; they concentrate upon what they can observe about the other as system-in-an-environment, as input and output, and they learn self-referentially, each within its own observational perspective. They can try to influence what they observe through their own action, and they can learn once again from feedback. In this way, an emergent order can arise … We call this … the social system." (Luhmann, 1984, at p. 157).

In other words: "what a burden is assumed by a theory that divides up linguistic structures that cover both the psychic and the social dimensions into two different systems" (Habermas, 1987, at p. 379).

Künzler (1987) already recognized that coding in Luhmann's theory is based on the model of DNA, that is, in terms of an ON/OFF scheme. Thus, the "closed systems" are shaped using a biological model. Unfortunately, Luhmann (1975a, 1984) used this biological metaphor instead of following Parsons' linguistic understanding of coding. In summary, Giddens (1984) and Habermas (1987) were right that Luhmann's theory had meta-biological overtones (Leydesdorff, 2006a and 2010b). Structural coupling can be considered as a biological mechanism: a network system is "plastic" with reference to the distribution of agents at the nodes firing. However, each agent in this case is counted only as ON/OFF, and hence not in terms of what the communication *means* for the communication as potentially different from the meanings individually provided to it.

The biological agents at the nodes have no choice other than ON/OFF in reaction to an update at the network level. The reflexive operations among psychological systems (*cogitantes*) and social systems (*cogitata*), however, evaluate the communications not only in terms of relative frequencies, but also substantively. *Cogitantes* can do this consciously, while the *cogitata* can be structured by the codes of communication and thus configurational meaning is provided to each communication of meaning.

In other words, the meanings are interfaced *twice*: as in the cybernetic model of Maturana & Varela developed for biological systems, the interactions among self-organizing subsystems have to be retained in a structural setting at each moment of time. Luhmann used the concept of "organization" for this level. Specific to this level is the decision-making about the boundary of meaning-processing at that moment of time. Note that this decision-making structure operates as a reflexive control mechanism on the self-organizing fluxes of communication (Achterberg & Vriens, 2009, at pp. 113 ff.). Stoyanova Kennedy & Kennedy (2010) identify this control mechanism as a "facilitator." From a communication-theoretical perspective, however, this mechanism does not have to function at the individual level, but can also operate more abstractly, for example, as a decision rule (Leydesdorff, 2006b, at pp. 139 ff.).

In addition to this first interface, the two *meaning-processing* systems—that is, consciousness and communication—have access to each other's substance reflexively, and hyper-reflexively across these reflections to one another. While the first coupling between consciousness and communication as different systems can be considered "structural" in the sense that the one system cannot operate without the other, their reflexive access to each other by this interpenetration remains operational since mediated (Luhmann, 1991; 2002, at pp. 175 and 182). In a first layer the two systems (*cogitantes* and *cogitata*) are constitutive of each other in a co-evolution; in a second layer, the systems' operations (that is, meaning processing) can be mediated across systems divides, for example, by language. The mediation serves the structuration of (uncertain!) expectations.

The operator for the communication of meaning is the evolutionary achievement of human language. Messages not only contain information, but can also be provided with meaning. We can understand one another, and—even more importantly—we can (hyper-) reflexively know a previous understanding to have been a misunderstanding. Particularly, the latter process of cognitive learning distinguishes human beings from Maturana's (1980; 1993) animal kingdom. Thus, one can begin to understand, within the perspective of social systems theory, the role of facilitation and learning that in principle enables us to improve and extend our communicative competencies and thus to prepare for and reflect on the new options that emerge in the self-organization of the communication of meaning.

**Conclusions and summary**

I have wished to argue that the ideological stalemate between (structure-oriented) systems theory and (action-oriented) critical philosophy can be circumvented from the perspective of communication theory. Both the vocabulary of *autopoiesis* and that of communicative action can enrich the semantics of such theorizing. The interhuman communication systems are not closed, but may tend to be so; for example, in restricted discourse or in scientific discussions. However, these non-linear dynamics emerge on the basis of interactions in which both uncertainty and meaning can be communicated. Uncertainty is continuously generated. Meaning emerges by positioning and relating bits of information in configurations. This can be measured using, for example, semantic maps.

Meanings can further be codified. Knowledge emerges recursively from relating different meanings. Knowledge perhaps can be considered as "a meaning that makes a difference" (cf. Bateson, 1972, at p. 453). At the network level of interpersonal communication, one can entertain discursive knowledge as different from personal reflections. Discursive knowledge, however, is no longer structured from below, but in terms of a specific code of communication that functions at the symbolic level. This code is reproduced and potentially changed by the participants in the communication because they have access to the substance of the communication, and this very access may induce learning and eventually a paradigm change. For example, when the code is no longer sufficient for processing the complexity in the communication, the communication system may go into crisis and new codes may be developed.

This process can be supported at the organizational level by facilitating the appreciation of what is in and out, and why! The specification of the "why" in the model brings us back to Habermas and Giddens; from this perspective, one can expect a social system to be *quasi-autopoietic* (Collier, 2008). The discussion of the "why" enables us to improve or worsen the communication reflexively. However, the inquiry is performed in terms of communications using codes of communication. Insofar as the constructs (e.g., codified knowledge) feed back on the constructors, Stoyanova Kennedy & Kennedy's (2010) communities of inquiry can increasingly be considered as a dependent variable of the inquiring communications.

A community of inquiry can be considered as an instantiation of the inquiring communication structures that develop using codes for focusing the communications. One is reflexively able to hypothesize the codes relevant to a discourse. This reflection spans another discourse. The reflection by the facilitator on the discourse—whether an individual or institutional agency—can transform the dynamics of communication by delineating the organization in terms of interfacing these hypothesized dimensions. The dynamics of communication, however, can be expected to feed back on the dynamics of the inquiring community to the extent that the latter may also be dissolved. I agree that these processes are open and thus a possible subject of empirical investigation. From this sociological perspective, observations may enable us to test heuristically specified expectations.